\documentstyle[12pt]{article}                    
\newfont{\ffont}{msym10}                          
\newcommand{\beq}{\begin{equation}}               
\newcommand{\eeq}{\end{equation}}                 
\newcommand{\bqry}{\begin{eqnarray}}              
\newcommand{\eqry}{\end{eqnarray}}                
\newcommand{\bqryn}{\begin{eqnarray*}}            
\newcommand{\eqryn}{\end{eqnarray*}}              
\newcommand{\preprint}[1]{\begin{table}[t]        
            \begin{flushright}                    
            \begin{large}{#1}\end{large}          
            \end{flushright}                      
            \end{table}}                          
\newcommand{\PD}[2]                               
    {\frac{\partial^{#2}}{\partial #1^{#2}}}      
\begin{document}
\preprint{LA-UR-97-XXXX}
\title{Constraint on Axial-Vector Meson Mixing Angle from Nonrelativistic
Constituent \\ Quark Model}
\author{\\ L. Burakovsky\thanks{E-mail: BURAKOV@PION.LANL.GOV} \
and \ T. Goldman\thanks{E-mail: GOLDMAN@T5.LANL.GOV} \
\\  \\  Theoretical Division, MS B285 \\  Los Alamos National Laboratory \\ 
Los Alamos, NM 87545, USA \\}
\date{ }
\maketitle
\begin{abstract}
In a nonrelativistic constituent quark model we find a constraint on the mixing
angle of the strange axial-vector mesons, $35^o\stackrel{<}{\sim }\theta _K
\stackrel{<}{\sim }55^o,$ determined solely by two parameters: the mass 
difference of the $a_1$ and $b_1$ mesons and the ratio of the constituent 
quark masses.
\end{abstract}

\bigskip
\section{Introduction}
It is known that the decay of the $I=1/2$ 1 $^3P_1$ and 1 $^1P_1$ mesons,
$K_1(1270)$ and $K_1(1400),$ with masses $1273\pm 7$ MeV and $1402\pm 7$ MeV, 
respectively \cite{pdg}, satisfies a dynamical selection rule such that
$$\Gamma \left( K_1(1270)\rightarrow K\rho \right) >> \Gamma \left( K_1(1270)
\rightarrow K^\ast \pi \right) ,$$
$$\Gamma \left( K_1(1400)\rightarrow K^\ast \pi \right) >> \Gamma \left( K_1(
1400)\rightarrow K\rho \right) ,$$
which, following the classical example of neutral kaons, suggests a large 
mixing (with a mixing angle close to $45^o)$ between the $I=1/2$ members of 
two axial-vector and nonets, $K_{1A}$ and $K_{1B},$ respectively, leading to 
the physical $K_1$ and $K_1^{'}$ states \cite{K1}. Carnegie {\it et al.}
\cite{Car} obtained the mixing angle $\theta _K=(41\pm 4)^o$ as the 
optimum fit to the data as of 1977. In a recent paper by Blundell {\it et al.}
\cite{BGP}, who have calculated strong OZI-allowed decays in the 
pseudoscalar emission model and the flux-tube breaking model, the $K_{1A}$-$K_{
1B}$ mixing angle obtained is $\simeq 45^o.$ Theoretically, in the exact 
$SU(3)$ limit the $K_{1A}$ and $K_{1B}$ states do not mix, similarly to their 
$I=1$ counterparts $a_1$ and $b_1.$ As for the $s$-quark mass greater than the 
$u$- and $d$-quark masses, $SU(3)$ is broken and these states do mix to give 
the physical $K_1$ and $K_1^{'}.$ If the $K_{1A}$ and $K_{1B}$ are degenerate 
before mixing, the mixing angle will always be $\theta _K=45^o$ \cite{CR,Lip}.
As pointed out by Suzuki \cite{Suz}, the data on $K\pi \pi $ production
in $\tau $-decay may confirm or refute this simple picture: if $\theta _K=45^
o,$ production of the $K_1(1270)$ and $K_1(1400)$ would be one-to-one up to 
the kinematic corrections, since in the $SU(3)$ limit only the linear 
combination $\left( K_1(1270)+K_1(1400)\right) /\sqrt{2}$ would have the right
quantum numbers to be produced there. After phase-space correction, the 
$K_1(1270)$ production would be favored over the $K_1(1400)$ one by nearly a 
factor of 2. However, current experimental data are very uncertain. The 
measurements made by the TPC/Two-Gamma collaboration give \cite{TPC}
\bqry
B\left( \tau \rightarrow \nu K_1(1270)\right)  & = & \left( 0.41^{+0.41}_{
-0.35}\right) \times 10^{-2}, \\
B\left( \tau \rightarrow \nu K_1(1400)\right)  & = & \left( 0.76^{+0.40}_{
-0.33}\right) \times 10^{-2}, \\
B\left( \tau \rightarrow \nu K_1\right)  & = & \left( 1.17^{+0.41}_{-0.37}
\right) \times 10^{-2}.
\eqry
Alemany \cite{Al} combines the CLEO and ALEPH data \cite{CA} to obtain
\beq
B\left( \tau \rightarrow \nu K_1\right) =(0.77\pm 0.12)\times 10^{-2},
\eeq
which is smaller, but consistent with, the TPC/Two-Gamma Results. Conversely,
the claim from the CLEO collaboration is that the $\tau $ decays 
preferentially into the $K_1(1270).$ If one assumes, however, that the 
production of the $K_1(1400)$ is favored over that of $K_1(1270)$ by nearly a 
factor of 2 (as follows from (1),(2) if the experimental errors are ignored), 
one would arrive at $\theta _K\approx 33^o$ \cite{Suz}. A very recent analysis
by Suzuki of the experimental data on the two-body decays of the $J/\psi $ and
$\psi ^{'}$ into an axial-vector and a pseudoscalar mesons from the BES 
collaboration \cite{BES} shows that any value of $\theta _K$ between $30^o$ 
and $60^o$ can be consistent with the $1^{+}0^{-}$ modes of both the $J/\psi $
and $\psi ^{'}$ that have been so far measured \cite{Suz1}. 

The purpose of this work is to consider the $K_{1A}-K_{1B}$ mixing within the
framework of a constituent quark model. In our previous papers \cite{P,D} this
model was successfully applied to $P$- and $D$-wave meson spectroscopy in 
order to explain the common mass near-degeneracy of two pairs of nonets, 
$(1\;^3P_0,\;1\;^3P_2),$ $(1\;^3D_1,\;1\;^3D_3),$ in the isovector and 
isodoublet channels, as observed in experiment, and to make predictions 
regarding the masses of missing and problematic $q\bar{q}$ states. As we shall
see, the nonrelativistic constituent quark model provides a very simple 
constraint on the $K_{1A}-K_{1B}$ mixing angle determined solely by the mass 
difference of the isovector counterparts of the corresponding nonets, the 
$a_1$ and $b_1$ mesons, and the ratio of the constituent quark masses. 

\section{Nonrelativistic constituent quark model}
In the constituent quark model, conventional mesons are bound states of a spin
1/2 quark and spin 1/2 antiquark bound by a phenomenological potential which 
has some basis in QCD \cite{LSG}. The quark and antiquark spins combine to 
give a total spin 0 or 1 which is coupled to the orbital angular momentum $L.$
This leads to meson parity and charge conjugation given by $P=(-1)^{L+1}$ and 
$C=(-1)^{L+S},$ respectively. One typically assumes that the $q\bar{q}$ wave 
function is a solution of a nonrelativistic Schr\"{o}dinger equation with the 
generalized Breit-Fermi Hamiltonian\footnote{The most widely used potential 
models are the relativized model of Godfrey and Isgur \cite{GI} for the 
$q\bar{q}$ mesons, and Capstick and Isgur \cite{CI} for the $qqq$ baryons. 
These models differ from the nonrelativistic quark potential model only in
relatively minor ways, such as the use of $H_{kin}=\sqrt{m_1^2+{\bf p}_1^2}+
\sqrt{m_2^2+{\bf p}_2^2}$ in place of that given in (5), the retention of the 
$m/E$ factors in the matrix elements, and the introduction of coordinate
smearing in the singular terms such as $\delta ({\bf r}).$}, $H_{BF},$
\beq
H_{BF}\;\psi _n({\bf r})\equiv \left( H_{kin}+V({\bf p},{\bf r})\right) \psi _
n({\bf r})=E_n\psi _n({\bf r}),
\eeq
where $H_{kin}=m_1+m_2+{\bf p}^2/2\mu -(1/m_1^3+1/m_2^3){\bf p}^4/8,$ $\mu =m_
1m_2/(m_1+m_2),$ $m_1$ and $m_2$ are the constituent quark masses, and to
first order in $(v/c)^2={\bf p}^2c^2/E^2\simeq {\bf p}^2/m^2c^2,$ $V({\bf p},
{\bf r})$ reduces to the standard nonrelativistic result, 
\beq
V({\bf p},{\bf r})\simeq V(r)+V_{SS}+V_{LS}+V_T,
\eeq
with $V(r)=V_V(r)+V_S(r)$ being the confining potential which consists of a
vector and a scalar contribution, and $V_{SS},V_{LS}$ and $V_T$ the spin-spin,
spin-orbit and tensor terms, respectively, given by
\cite{LSG}
\beq
V_{SS}=\frac{2}{3m_1m_2}\;{\bf s}_1\cdot {\bf s}_2\;\triangle V_V(r),
\eeq
$$V_{LS}=\frac{1}{4m_1^2m_2^2}\frac{1}{r}\left( \left\{ [(m_1+m_2)^2+2m_1m_2]\;
{\bf L}\cdot {\bf S}_{+}+(m_2^2-m_1^2)\;{\bf L}\cdot {\bf S}_{-}\right\}
\frac{dV_V(r)}{dr}\right. $$
\beq
\left. -\;[(m_1^2+m_2^2)\;{\bf L}\cdot {\bf S}_{+}+(m_2^2-m_1^2)\;{\bf L}\cdot
{\bf S}_{-}]\;\frac{dV_S(r)}{dr}\right) ,
\eeq
\beq
V_T=\frac{1}{12m_1m_2}\left( \frac{1}{r}\frac{dV_V(r)}{dr}-\frac{d^2V_V(r)}{
dr^2}\right) S_{12}.
\eeq
Here ${\bf S}_{+}\equiv {\bf s}_1+{\bf s}_2,$ ${\bf S}_{-}\equiv {\bf s}_1-
{\bf s}_2,$ and
\beq
S_{12}\equiv 3\left( \frac{({\bf s}_1\cdot {\bf r})({\bf s}_2\cdot {\bf r})}{
r^2}-\frac{1}{3}{\bf s}_1\cdot {\bf s}_2\right).
\eeq
For constituents with spin $s_1=s_2=1/2,$ $S_{12}$ may be rewritten in the form
\beq
S_{12}=2\left( 3\frac{({\bf S}\cdot {\bf r})^2}{r^2}-{\bf S}^2\right),\;\;\;
{\bf S}={\bf S}_{+}\equiv {\bf s}_1+{\bf s}_2.
\eeq
Since $(m_1+m_2)^2+2m_1m_2=6m_1m_2+(m_2-m_1)^2,$ $m_1^2+m_2^2=2m_1m_2+(m_2-m_
1)^2,$ the expression for $V_{LS},$ Eq. (8), may be rewritten as
$$V_{LS}=\frac{1}{2m_1m_2}\frac{1}{r}\left[ \left( 3\frac{dV_V(r)}{dr}-
\frac{dV_S(r)}{dr}\right) + \frac{(m_2-m_1)^2}{2m_1m_2}\left(\frac{dV_V(r)}{d
r}-\frac{dV_S(r)}{dr}\right) \right] {\bf L}\cdot {\bf S}_{+}$$
\beq
+\frac{m_2^2-m_1^2}{4m_1^2m_2^2}\;\frac{1}{r}\left( \frac{dV_V(r)}{dr}-\frac{
dV_S(r)}{dr}\right) {\bf L}\cdot {\bf S}_{-}\equiv V_{LS}^{+}+V_{LS}^{-}.
\eeq
Since two terms corresponding to the derivatives of the potentials with respect
to $r$ are of the same order of magnitude, the above expression for 
$V_{LS}^{+}$ may be rewritten as
\beq
V_{LS}^{+}=\frac{1}{2m_1m_2}\frac{1}{r}\left( 3\frac{dV_V(r)}{dr}-\frac{dV_
S(r)}{dr}\right) {\bf L}\cdot {\bf S}\left[ 1+\frac{(m_2-m_1)^2}{2m_1m_2}\;O(
1)\right] .
\eeq

\section{$P$-wave meson spectroscopy}
We now wish to apply the Breit-Fermi Hamiltonian to the $P$-wave mesons. By
calculating the expectation values of different terms of the Hamiltonian 
defined in Eqs. (7),(11),(12), taking into account the corresponding matrix 
elements $\langle {\bf L}\cdot {\bf S}\rangle $ and $S_{12}$ \cite{LSG}, one 
obtains the relations \cite{BGP,P}
\bqryn
M(^3P_0) & = & M_0+\frac{1}{4}\langle V_{SS}\rangle -2\langle V_{LS}^{+}
\rangle - \langle V_T\rangle , \\
M(^3P_2) & = & M_0+\frac{1}{4}\langle V_{SS}\rangle +\langle V_{LS}^{+}\rangle
- \frac{1}{10}\langle V_T\rangle , \\  
M(a_1) & = & M_0+\frac{1}{4}\langle V_{SS}\rangle -\langle V_{LS}^{+}\rangle +
\frac{1}{2}\langle V_T\rangle , \\
M(b_1) & = & M_0-\frac{3}{4}\langle V_{SS}\rangle ,
\eqryn 
$$\left( \begin{array}{c}
M(K_1) \\ M(K_1^{'}) \end{array} \right) =\left( \begin{array}{cc}
M_0+\frac{1}{4}\langle V_{SS}\rangle -\langle V_{LS}^{+}\rangle +\frac{1}{2}
\langle V_T\rangle  & \sqrt{2}\langle V_{LS}^{-}\rangle \\ 
\sqrt{2}\langle V_{LS}^{-}\rangle  & M_0-\frac{3}{4}\langle V_{SS}\rangle 
\end{array} \right) \left( \begin{array}{c}
K_{1A} \\ K_{1B} \end{array} \right) ,$$
where $M_0$ stands for the sum of the constituent quark masses and binding 
energies in either case. The $V_{LS}^{-}$ term acts only on the $I=1/2$ 
singlet and triplet states giving rise to the spin-orbit mixing between these 
states\footnote{The spin-orbit $^3P_1-^1P_1$ mixing is a property of the model
we are considering; the possibility that another mechanism is responsible for 
this mixing, such as mixing via common decay channels \cite{Lip} should not be
ruled out, but is not included here.}, and is responsible for the physical 
masses of the $K_1$ and $K_1^{'}.$ The masses of the $K_{1A}$ and $K_{1B}$
are determined by relations which are common for all eight $I=1,1/2$ $P$-wave 
mesons, $b_1,a_0,a_1,a_2,K_{1B},K_0^\ast ,K_{1A},K_2^\ast :$
\bqry
M(^1P_1) & = & m_1+m_2+E_0-\frac{3}{4}\frac{a}{m_1m_2}, \\
M(^3P_0) & = & m_1+m_2+E_0+\frac{1}{4}\frac{a}{m_1m_2}-\frac{2b}{m_1m_2}-
\frac{c}{m_1m_2}, \\
M(^3P_1) & = & m_1+m_2+E_0+\frac{1}{4}\frac{a}{m_1m_2}-\frac{b}{m_1m_2}+
\frac{c}{2m_1m_2}, \\
M(^3P_2) & = & m_1+m_2+E_0+\frac{1}{4}\frac{a}{m_1m_2}+\frac{b}{m_1m_2}-
\frac{c}{10m_1m_2},
\eqry 
where $a,b$ and $c$ are related to the matrix elements of $V_{SS},$ $V_{LS}$ 
and $V_T$ (see Eqs. (7),(9),(13)), and assumed to be the same for all of the 
$P$-wave states. $E_0$ is a nonrelativistic binding energy which may in 
general be absorbed in the definition of a constituent quark mass \cite{P,D}.
We assume also $SU(2)$ flavor symmetry: $m(u)=m(d)\equiv n,$ $m(s)\equiv s.$

The correction to $V_{LS}^{+}$ in the formula (13), due to the difference in 
the masses of the $n$ and $s$ quarks, is ignored. Indeed, these effective 
masses, as calculated from (14)-(17) in the case where $E_0$ is absorbed into
their definition, are\footnote{In the following, $a_0$ stands for the mass of 
the $a_0,$ etc.}
\beq
n=\frac{3b_1+a_0+3a_1+5a_2}{24},
\eeq
\beq
s=\frac{6K_{1B}+2K_0^\ast +6K_{1A}+10K_2^\ast -3b_1-a_0-3a_1-5a_2}{24}.
\eeq
With the physical values of the meson masses (in GeV), $a_1\simeq b_1\cong 
1.23,$ $a_0\simeq a_2\cong 1.32,$ $K_{1A}\simeq K_{1B}\cong 1.34,$ $K_0^\ast
\simeq K_2^\ast \cong 1.43,$ the above relations give 
\beq
n\simeq 640\;{\rm MeV,}\;\;\;s\simeq 740\;{\rm MeV,}
\eeq
so that the abovementioned correction, according to (13), is $\sim 100^2/(2
\cdot 640\cdot 740)\simeq 1$\%, i.e., comparable to isospin breaking on the 
scale considered here, and so completely negligible.
 
It follows from (14)-(17) that
\bqry
\frac{9a}{m_1m_2} & = & M(^3P_0)+3M(^3P_1)+5M(^3P_2)-9M(^1P_1), \\ 
\frac{12b}{m_1m_2} & = & 5M(^3P_2)-3M(^3P_1)-2M(^3P_0), \\ 
\frac{18c}{5m_1m_2} & = & 3M(^3P_1)-2M(^3P_0)-M(^3P_2). 
\eqry
By expressing the ratio $n/s$ in three different ways, viz., dividing the 
expressions (21)-(23) for the $I=1/2$ and $I=1$ mesons by each other, one 
obtains the relations
\beq
\frac{n}{s}=\frac{K_0^\ast +3K_{1A}+5K_2^\ast -9K_{1B}}{a_0+3a_1+5a_2-9b_1}=
\frac{5K_2^\ast -3K_{1A}-2K_0^\ast }{5a_2-3a_1-2a_0}=\frac{2K_0^\ast +K_2^
\ast -3K_{1A}}{2a_0+a_2-3a_1}.
\eeq
It follows from the last relation of (24) that
\beq
(K_2^\ast -K_0^\ast )(a_2-a_1)=(K_2^\ast -K_{1A})(a_2-a_0).
\eeq
This formula explains the common mass degeneracy of the scalar and tensor 
meson nonets in the isovector and isodoublet channels. Using now (24) and 
(25), one arrives, by straightforward algebra, at
\beq
\frac{n}{s}=\frac{K_{1A}-K_{1B}}{a_1-b_1}.
\eeq
This relation is an intrinsic property of the model we are considering; it 
depends neither on the values of the input parameters, $n,s,a,b,c,$ nor the
presence of $E_0$ in the relations (14)-(17). We shall now use this relation
in order to obtain a constraint on the $K_{1A}-K_{1B}$ mixing angle.
 
\section{Constraint on the $K_{1A}-K_{1B}$ mixing angle}
Since, on general grounds, $n\leq s,$ it follows from (26) that
\beq
\Big| K_{1A}-K_{1B}\Big| \leq \Big| a_1-b_1\Big| \equiv \Delta ,
\eeq
which may be rewritten as
\beq
K_{1A}^2+K_{1B}^2-2K_{1A}K_{1B}\leq \Delta ^2.
\eeq
Moreover, independent of the mixing angle,
\beq
K_{1A}^2+K_{1B}^2=K_1^2+K_1^{'2}.
\eeq
It then follows from (28),(29) that
\beq
2K_{1A}K_{1B}\geq K_1^2+K_1^{'2}-\Delta ^2.
\eeq
To obtain a constraint on the $K_{1A}-K_{1B}$ mixing angle, we now use the 
formula \cite{Suz}
$$\tan ^2(2\theta _K)=\left( \frac{K_1^2-K_1^{'2}}{K_{1B}^2-K_{1A}^2}
\right) ^2-1,$$ which may be rewritten as
\beq
\cos ^2(2\theta _K)=\left( \frac{K_{1B}^2-K_{1A}^2}{K_1^2-K_1^{'2}}\right) ^2.
\eeq
It follows from (29),(30) that
$$\left( K_{1B}^2-K_{1A}^2\right) ^2=\left( K_{1A}^2+K_{1B}^2\right) ^2-
4K_{1A}^2K_{1B}^2$$
\beq
\leq \left( K_1^2+K_1^{'2}\right) ^2-\left(K_1^2+K_1^{'2}-
\Delta ^2\right) ^2\cong 2\Delta ^2\left( K_1^2+K_1^{'2}\right) ,
\eeq
since $\Delta \sim 50$ MeV (see below), and therefore $\Delta ^2<<K_1^2+K_1^{
'2}.$ Thus, Eq. (31) finally reduces to
\beq
\cos ^2(2\theta _K)\leq \frac{2\Delta ^2(K_1^2+K_1^{'2})}{(K_1^2-K_1^{'2})^2},
\eeq
and therefore
\beq 
\Big| \cos (2\theta _K)\Big| \leq \frac{\Delta \sqrt{2(K_1^2+K_1^{'2})}}{
|K_1^2-K_1^{'2}|}.
\eeq
The value of $\Delta $ is determined by current experimental data on the $a_1$
and $b_1$ meson masses \cite{pdg}: $a_1=1230\pm 40$ MeV, $b_1=1231\pm 10$ MeV.
Therefore, $\Delta \leq 50$ MeV, and one obtains, from (34),
\beq
33.6^o\leq \theta _K\leq 56.4^o,
\eeq
consistent with the recent result of Suzuki \cite{Suz1}, $30^o\leq \theta _K
\leq 60^o.$ The above constraint may be tightened further by using the ratio 
of the constituent quark masses given in (21). Then from (26) we obtain
\beq
\Big| K_{1A}-K_{1B}\Big| =\frac{n}{s}\;\Big| a_1-b_1\Big| \leq \frac{0.64}{
0.74}\;50\;{\rm MeV}\simeq 43\;{\rm MeV}\equiv \Delta ^{'}.
\eeq
With this $\Delta {'}$ being used in (34) in place of $\Delta ,$ one obtains
\beq
35.3^o\leq \theta _K\leq 54.7^o.
\eeq
Both the ranges (35) and (37) are consistent with the value $\theta _K=(37.3
\pm 3.2)^o$ obtained in our previous work \cite{P}.

\section{Concluding remarks}
As we have shown, a nonrelativistic constituent quark model provides a simple
constraint on the $K_{1A}-K_{1B}$ mixing angle, in terms of the mass difference
of the $a_1$ and $b_1$ mesons and the squared masses of the physical states
$K_1$ and $K_1^{'}.$ The numerical value of the allowed interval for the mixing
angle, $33.6^o\leq \theta_K\leq 56.4^o,$ is consistent with that provided by
the very recent analysis by Suzuki \cite{Suz1}. This interval may be 
constrained further by using the ratio of the constituent quark masses. In
the mass degenerate case $a_1=b_1,$ the model considered shows a similar
mass degeneracy for the corresponding strange mesons, $K_{1A}=K_{1B},$ 
independent of the input parameters, and so requiring a precise $45^o$ mixing.
We conclude, therefore, that more precise experimental data on the mass of the
$a_1$ meson are required to obtain a better estimate of the $K_{1A}-K_{1B}$ 
mixing angle.

\end{document}